\documentclass[twocolumn,showpacs,floatfix,aps]{revtex4}
\usepackage{amssymb,amsmath}
\usepackage{subfigure}
\usepackage[dvips]{graphics,color}
\usepackage{epsfig}
\usepackage{float}
\newcommand{\bea}{\begin{array}}
\newcommand{\ear}{\end{array}}
\newcommand{\bege}{\begin{equation}}
\newcommand{\enge}{\end{equation}}

\newcommand{\beq}{\begin{eqnarray}}\newcommand{\benu}{\begin{enumerate}}\newcommand{\enu}{\end{enumerate}}
\newcommand{\eeq}{\end{eqnarray}}

\newcommand{\me}{\frac{1}{2}}
\newcommand{\noi}{\noindent}

\newcommand{\lie}{\text{{\it\char'44}}}
\definecolor{Magenta}{named}{Magenta}
\definecolor{MidnightBlue}{named}{MidnightBlue}
\definecolor{Emerald}{named}{Emerald}
\definecolor{YellowOrange}{named}{YellowOrange}
\definecolor{Plum}{named}{Plum}
\newcommand{\be}{\beta}

\begin{document}

\title{Extra dimensions in LHC via mini-black holes: effective Kerr-Newman braneworld effects}

\author{R. da Rocha}
\email{roldao@ifi.unicamp.br}
\affiliation{Instituto de F\'{\i}sica Te\'orica, 
Universidade Estadual Paulista, 
Rua Pamplona 145\\
01405-900 S\~ao Paulo, SP, Brazil\\and\\
DRCC - Instituto de F\'{\i}sica Gleb Wataghin\\ Universidade Estadual de Campinas
CP 6165, 13083-970 Campinas, SP, Brazil}
\author{C. H. Coimbra-Ara\'ujo}
\affiliation{Instituto de F\'{\i}sica Gleb Wataghin, Universidade Estadual de Campinas,
CP 6165, 13083-970 Campinas, SP, Brazil.}
\email{carlosc@ifi.unicamp.br}


\begin{abstract}
 We solve Einstein equations 
on the brane to derive the exact form of the braneworld-corrected perturbations in Kerr-Newman singularities, using Randall-Sundrum and Arkani-Hamed-Dimopoulos-Dvali (ADD) models. 
It is a consequence of such models the possibility that Kerr-Newman mini-black holes can 
be produced in LHC. %, and for the case where there is no braneworld corrections, our
 %mini-black hole is a conventional Kerr black hole, with no bulk-induced tidal charge.  
We use this approach 
to derive a normalized correction for the Schwarzschild Myers-Perry radius of a static $(4+n)$-dimensional mini-black hole, using more 
realistic approaches arising from Kerr-Newman mini-black hole analysis. 
Besides, we prove that there are four  Kerr-Newman black hole horizons in braneworld scenario we use, 
although only the outer horizon is relevant in the physical measurable processes. Parton cross sections in LHC and Hawking temperature are also investigated 
as functions of Planck mass (in the LHC range 1-10 TeV), mini-black hole mass and the number of large extra dimensions in
 braneworld large extra-dimensional scenarios. 
In this case a more realistic brane effect-corrected formalism can achieve more precisely 
the effective extra-dimensional Planck mass and the number of large 
extra dimensions  --- in  Arkani-Hamed-Dimopoulos-Dvali model ---
or the size of the warped  extra dimension --- in Randall-Sundrum formalism.
\end{abstract}
\maketitle
The hierarchy problem, concerning the large ratio of the weak scale mass and the Planck scale mass, can possibly shed new light on the 
existence of large extra dimensions   and, in order to improve 
some theories that have 
not been tested yet, and their predictions --- like  compactification of extra 
dimensions of the order of the Planck length ($\approx$ 10$^{-33}$ cm) --- 
there has recently been proposed new approaches \cite{hie2,Randall2,Randall,todos,todos1} 
suggesting that extra dimensions can be as large as around a millimeter. One of these approaches, 
Arkani-Hamed-Dimopoulos-Dvali (ADD) formalism, essentially suggests a higher-dimensional compact manifold to describe the universe, and 
explains how large extra dimensions could dilute the strength of gravity \cite{todos}. 
Although the gravitational force would be very strong in higher-dimensional spacetime, in the lower-dimensional effective theory
gravity is weaker when the volume of the extra-dimensional compactified space is bigger. ADD predicted that this dilution of gravity
into extra dimensions could possibly be so large that it can explain why the observed 4-dimensional gravity is so feeble.
ADD models
all have necessarily more than one extra curled up dimension -- in order not to imply deviations in Newtonian gravity over solar system 
distances ---  wherein a single brane accommodates Standard Model particles.
ADD models propose that the fundamental mass scale determining gravity strength is close to a TeV, instead 
of the Planck scale mass. It represents the
fundamental higher-dimensional Planck scale. 

Another formalism concerning large extra dimensions is the Randall-Sundrum  braneworld model \cite{Randall2,Randall}.
This formalism is 
effectively 5-dimensional, with one warped extra dimension,
  and it is based on a 5-dimensional reduction of (1+9)-dimensional brane Horava-Witten theory
\cite{hw,luka1,luka2,luka3}.  
The idea that the universe 
is trapped on a membrane in some higher-dimensional spacetime may explain why gravity is so weak, and could be
 tested at high energy particle accelerators. In this context, the observable universe proposed by Randall and Sundrum, 
in one of their two models, can be described as a  
brane embedded in an AdS$_5$ bulk. 
At low energies gravity is localized on the brane and general relativity is recovered, but, at high energies,
 significant changes are introduced in gravitational dynamics, forcing general relativity to break down and 
to be overcome by a quantum gravity theory \cite{rov,casa2}.
A plausible reason for the gravitational force to appear to be so weak can be its dilution in possibly existing extra dimensions
related to a bulk, where $p$-branes \cite{gr, zwi, zwi1, zwi2, bs, Townsend} are embedded. $p$-branes are good candidates for braneworlds because they possess gauge symmetries \cite{zwi, zwi1, zwi2} and automatically incorporate a quantum theory of gravity. 
%By supposing fermion and Standard Model gauge fields reside in the 3-brane 
%and only gravity can smear out extra dimensions, the weakness of gravity in comparison with other fundamental forces could be explained. 
%Some experiments involving the direct measurement of Newtonian gravity put limits on the size of extra dimensions to a value of less than a few tenths of 
%a millimeter. Using such an approach, the traditional Planck energy $\approx$ 10$^{19}$ GeV
%is no more than an effective scale and the real $D$-dimensional fundamental Planck scale is deeply distinct from 4-dimensional Planck scale. 
%For $D=10$ and radii associated with the extra dimensions of the Fermi scale, 
%we obtain Planck energy of order of TeVs. http://www.google.com.br/
%An alternative scenario can be achieved using effective 5-dimensional theories over manifolds endowed with
%the Randall-Sundrum  metric (RS), which induces a volcano barrier-shaped effective potential for gravitons around the brane \cite{Likken}.
%The corresponding spectrum of gravitational perturbations has a massless bound state on the brane,
%and a continuum of bulk modes with suppressed couplings to brane fields. These bulk modes introduce small
%corrections at short distances, and the introduction of more compact dimensions does not affect the localization of matter fields.
%However, true localization takes place only for massless fields \cite{Gregory}, and in the massive case the bound state becomes metastable,
%being able to leak into the extra dimension(s). 

In Randall-Sundrum formalism gravity is also diluted in warped extra-dimensional space, but this time due to graviton's probability 
function (GPF) that is dependent  on the distance to the gravity brane of the theory. 
While in ADD models the GPF is equally distributed over all the extension
of extra-dimensional space, and 4-dimensional gravity becomes diluted, in Randall-Sundrum formalism GPF varies due to the warped geometry.
By the effective warp-factor in Randall-Sundrum formalism, all fundamental masses, expected to be of the order of $M_P$ (the Planck 
scale mass) on an auxiliary (gravity) brane, become $10^{16}$ smaller, about a TeV magnitude. All masses are rescaled by the 
warp-factor, and the 5-dimensional Planck
scale is $10^{16}$ TeV, whereas the $M_P$ denotes now the effective warp-factor suppressed
Planck scale.

In previous papers we have explored Randall-Sundrum formalism 
in the context of supermassive black holes (BHs)  present in the nucleus of galaxies and quasars,
and when they cause deviations from the $4D$ general relativity,  these corrections should cause a small deviation in all BH properties \cite{rcp,jcap}. It has been already proposed the observation of braneworld BHs, and 
there are also other proposals on probing extra dimensions \cite{ruth33,sak}. Now 
we want to investigate these braneworld effects for the case of mini-BHs and their possible measurable 
consequences in Large Hadron Collider (LHC) observations \cite{algre}. It is claimed that if the fundamental Planck scale is a TeV scale,  LHC will produce over $10^7$ black holes per year \cite{dim1}.

Mini-BHs have radii
 that are much smaller 
than the size of 
extra dimensions, and therefore can be considered as 
totally embedded in a $D$-dimensional space. It is argued that as particles approach each other in LHC, their gravitational attraction
 increases steadily, and they can enter extra-dimensional space when they are extremely close. In that case it 
would allow gravity to increase and a mini-BH could form.
The mass of a mini-BH can be no smaller than Planck mass, of order 22 micrograms, corresponding to 
 mini-BH radius of the quantum fuzzy foam ($10^{-33}$ m).
Also, mini-BH lifetime can be about $10^{-26}$ s and its temperature, typically around 80 GeV $\approx$ $1.5 \times 10^{14}$ K, 
is much lower than it would be in a 4-dimensional spacetime, but still  
presents a high surface temperature though, at which mini-BHs would evaporate very rapidly into 
 photons, electrons, and quarks, with energies ranging from 80 GeV down.
It is believed that mini-BHs would either evaporate completely or
leave a remnant, and lifetime estimates were obtained in \cite{casa1}. Although 
the Standard Model does not put a constraint on the minimum size BH production by an accelerator, of 
about less than 10$^{16}$ TeV, some astounding ideas suggest that gravity becomes stronger at 
small distances because of the effects of extra dimensions accessed only
 by gravity. 

First of all the reconstruction of temperature 
as a function of mini-BH mass provides knowledge concerning 
the extra dimensions of spacetime \cite{dim1,land,land1}. In the case of Planck scales close to  TeV, the number of extra
 dimensions could thus be  easily unraveled by the features of emitted particles. The mini-BHs production in LHC \cite{bar2,gid} 
is an expected possibility and, as it is claimed,
 this will be the dominant effect in LHC under a quantum gravity theory  \cite{argyres,empa,banks}.
Also, a complete characterization of the Hawking radiation in  extra-dimensional scenarios, showing that graviton emission becomes relevant for a large number of extra
dimensions, can be seen in \cite{card1,card2,sto1,sto2,sto3}. 
%A mini-BH like those LHC might produce would have a very small radius, around 2 $\times 10^{-19}$ m. 
Exact stationary axisymmetric solutions describing rotating BHs --- with tidal
charge as well as with electric charge --- localized on the brane in
RS braneworld model have been investigated in \cite{tce}. Also, new solutions 
of the Einstein-Maxwell equations that describes an electrically charged and slowly rotating BH in five dimensions
was discussed in \cite{al1,al2,posic2006}.

We consider both Randall-Sundrum and ADD mini-BH production, because they present different character, 
as pointed out by Stojkovic \cite{sto3}. 
While ADD mini-BHs have the first phase Hawking radiation  mostly in
the bulk (and the second phase is mostly on the brane), existence of relaxation time during which the black
hole looses the bulk components of angular momentum, and recoil effect to leave the brane, a Randall-Sundrum mini-BH
have bulk radiation strongly suppressed, do not have any bulk components of angular momentum (absence of relaxation time)
and cannot recoil and leave the brane.

The main aim of this paper is to show how to correct, considering braneworld effects in ADD and Randall-Sundrum models, 
the mini-BH Kerr-Newman radii (horizons), 
by solving Einstein equations on the brane. In the case we consider, when the radius of a BH on the
brane is much smaller than the size scale of the extra dimensions,
the BH can be well
described by the classical solutions of higher-dimensional
Einstein equations. Schwarzschild  radius, in the context of Myers-Perry extra-dimensional formalism, is shown here to be significantly increased  by tidal charge and spinning effects.  The corrections we obtain give rise to
more precise calculations concerning cross sections, Planck and mini-BHs masses, and Hawking temperature,
 contributing in this way 
to a more complete, precise and realistic analysis of mini-BH production in the next generation of particle colliders, 
such as LHC. 

This article  is organized as follows: in Section I Einstein field equations on the brane are presented together 
with the brane-corrected  Newtonian potential. Reissner-Nordstr\o m BHs are also considered in a braneworld scenario
viewpoint. 
In Section II by inputting a radial coordinate-valued function to perturb a Kerr-Newman singularity,
we  solve Einstein field equations on the brane to find the explicit form of these perturbations.
Various graphics are depicted to illustrate the corrections in Kerr-Newman radii as functions of the
BH angular momentum, both in Randall-Sundrum and ADD  braneworld scenarios, for various fixed values of the BH charge. 
Also, cross sections are now corrected by braneworld effects 
and it gives rise to more realistic analysis of LHC collisions products, such as mini-BH and particle production via mini-BH decay.
The corrections in Kerr-Newman mini-BH radii are analyzed separately in a Randall-Sundrum braneworld scenario --- where there is $n=1$
 warped extra dimension only --- and in ADD scenario with $n= 2,4$ and 6 extra dimensions,
 as functions of Planck masses in the range 1-10 TeV, as well as BH masses and the BH spinning parameter.  
  Graphics of parton cross sections are also shown as functions of Planck masses in the range 1-10 TeV and BH masses. Finally in Section III
Kerr-Newman mini-BH Hawking evaporation and temperature are investigated in a braneworld scenario. All graphics present 
our results in explicit comparison with Schwarzschild BH described by Myers-Perry standard results --- see, e.g., \cite{dim1,land,land1}.

\section{Mini-black holes on the brane}

In a braneworld scenario the Einstein field equations read \cite{Shiromizu,Maartens,rcp,jcap}
\begin{eqnarray}\label{123}
&&G_{\mu\nu} = -\frac{1}{2}{\Lambda}_5g_{\mu\nu}\nonumber\\
&&+ \frac{1}{4}\kappa_5^4\left[TT_{\mu\nu} - T^{\;\,\alpha} _{ \nu}T_{\mu \alpha} + \frac{1}{2}g_{\mu\nu}(T^2 - T_{\alpha\beta}^{\;\;\;\;\alpha\beta})\right] - E_{\mu\nu},\nonumber
\end{eqnarray}
\noindent where $T = T_\alpha^{\;\,\alpha}$ denotes the trace of the energy-momentum tensor, $\Lambda_5$ denotes 
the 5-dimensional cosmological bulk constant, and  $E_{\mu\nu}$ denotes the `electric' components of the Weyl tensor, which
 can be expressed by means of the
extrinsic curvature components $K_{\mu\nu} = -\displaystyle \frac{1}{2} \lie_n g_{\mu\nu}$ by \cite{soda}
\begin{equation}
E_{\mu\nu} = \lie_n K_{\mu\nu} + K_{\mu}^{\;\;\alpha}K_{\alpha\nu} - \frac{1}{\ell^2}g_{\mu\nu},
\end{equation}\noindent where $\ell$ denotes the bulk curvature radius. It corresponds equivalently 
to the effective size of the extra dimension probed by a 5$D$ graviton \cite{Randall,Randall2,Maartens}.
Let $\kappa_5$ be defined as  $\kappa_5 = 8\pi G_5$, where 
$G_5$ denotes the 5-dimensional Newton gravitational constant, which can be related to the
4-dimensional gravitational constant $G$ by $G_5 = G\ell_{\rm Planck}$, where $\ell_{\rm Planck} = \sqrt{G\hbar/c^3}$ is the Planck length.

As indicated in \cite{Randall,Maartens}, ``\ldots table-top tests of Newton's law currently find no deviations down to the order
of 0.1 mm\ldots'', so that $\ell \lesssim $ 0.1 mm \cite{empa,emparan}
 provides a more accurate magnitude limit improvement on the bulk curvature $\ell$. 
%by analyzing the existence of stellar-mass BHs
%on long time scales and of BH X-ray binaries. In this paper we relax the stringency $\ell \lesssim 0.01$ mm to the former table-top limit
%$\ell \lesssim $ 0.1 mm. 
The Weyl `electric' term $E_{\mu\nu}$  carries an imprint of high-energy effects sourcing KK modes. %It means that highly energetic
%stars and the process of gravitational collapse, and naturally  BHs, lead to deviations from the 4-dimensional general
%relativity problem. This occurs basically because the gravitational collapse unavoidably produces energies high enough to make
%these corrections significant. From the brane-observer viewpoint, the KK corrections in $E_{\mu\nu}$ are nonlocal, since they
%incorporate 5-dimensional gravity wave modes. These nonlocal corrections cannot be determined purely from data on the brane \cite{Maartens}.
The component $E_{\mu\nu}$ also carries information about the collapse process of BHs.
In the perturbative analysis of Randall-Sundrum positive tension 3-brane, KK modes consist of a continuous spectrum without any gap
 \cite{Gregory}. It
generates a correction in the gravitational potential $V(r) =\frac{GM}{c^2r}$ to 4$D$ gravity at low energies by
 extra-dimensional effects \cite{Maartens}, 
which is
given by \cite{Randall,Randall2}, for $r \gg \ell$
\begin{equation}\label{potential}
V(r) = \displaystyle\frac{GM}{c^2r}\left[1 + \frac{2\ell^2}{3r^2} + \mathcal{O}\left(\frac{\ell}{r}\right)^4\right],
\end{equation}
\noindent
and for $r \ll \ell$,
\begin{equation}\label{potential2}
V(r)  \approx  \displaystyle\frac{GM \ell}{c^2r^2}.
\end{equation}
The KK modes that generate these corrections are responsible for a nonzero $E_{\mu\nu}$. This term carries the modification to the weak-field field equations, as we have
already seen.

For a static spherical metric on the brane
given by \begin{equation}\label{124}
g_{\mu\nu}dx^{\mu}dx^{\nu} = - F(r)dt^2 + \frac{dr^2}{H(r)} + r^2d\Omega^2,
\end{equation}
\noindent
 the projected electric component Weyl term on the brane is given by the expressions \cite{Maartens,rcp,jcap}
\begin{eqnarray}\label{ewe}
E_{00} &=& \frac{F}{r}\left(H' - \frac{1 - H}{r}\right),\;\;
E_{rr} = -\frac{1}{rH}\left(\frac{F'}{F} - \frac{1 - H}{r}\right),\nonumber\\
 E_{\theta\theta} &=& -1 + H +\frac{r}{2}H\left(\frac{F'}{F} + \frac{H'}{H}\right).
\end{eqnarray}
\noindent Note that in Eq.(\ref{124}) the Schwarzschild metric is recovered if $F(r)$ equals $H(r)$. It is well known that 
the most general solutions of Eq.(\ref{124}), in the case of a Reissner-Nordstr\o m (RN) BH, are given by \cite{Maartens,Dadhich}
\begin{equation}\label{h}
F(r) =  1 - \frac{2GM}{c^2r} - \psi(r) = 1 - \frac{2GM}{c^2r} + \frac{2G\ell Q^*}{c^2r^2},
\end{equation}
\noindent
where $M$ and $Q^*$ denote respectively the mass and the tidal charge induced by the bulk of an effective RN 
BH. The bulk induces a tidal charge BH on the brane. The radial coordinate-valued $\psi(r)$ is obtained 
when one substitutes $F(r)$ in Einstein equations on the brane (see Eq.(\ref{ansa1})). We use in the next Section an analogous
procedure to obtain the braneworld form of Kerr-Newman metric.

As asserted in \cite{Maartens}, a negative $Q^*$ strengthens the gravi\-tational field, since it arises from the source mass $M$ on the brane. 
By contrast, in the RN solution of general relativity, $Q^*\sim q^2$, where $q$ denotes the RN mini-BH electric and tidal charges, 
and this weakens the gravitational field. Negative tidal charge also preserves 
the spacelike nature of the singularity. 
The tidal charge BH metric does not satisfy the far-field $r^{-3}$ correction to the gravitational potential, however Eq.(\ref{h}) 
shows the correct $5D$ behavior of the potential  at short distances, so that the tidal-charge metric could be
 a good approximation in the strong-field regime for small BHs \cite{Maartens}, as seen in Eq.(\ref{potential2}). 
Eqs.  (\ref{potential2}) and (\ref{h}) also imply, for extremal BHs defined by the relation
\begin{equation}\label{charge}
Q^* = -2M,
\end{equation}
that RN BHs 
in braneworld models also have two --- inner and outer --- horizons, denoted by $R_{{\rm RNbrane}}^\pm$, which can be obtained by fixing
 $F(r) = 0$ in  Eq.(\ref{h}). It results in \cite{Maartens}
%\begin{equation}
%1 - \frac{2GM}{c^2R_{{\rm RNbrane}}} + \frac{2G\ell Q^*}{c^2R_{{\rm RNbrane}}^2} = 0
%\end{equation}
%\noindent This equation can be rewritten as
%\begin{equation}
%R_{{\rm RNbrane}}^2 - \frac{2GM}{c^2}R_{{\rm RNbrane}} + \frac{2G\ell Q^*}{c^2} = 0, 
%\end{equation}
%\noindent
%with solution given by \cite{Maartens}
\bege\label{109}
R_{{\rm RNbrane}}^\pm = \frac{GM}{c^2} \pm \frac{1}{c}\left[\frac{G^2M^2}{c^2} - 2\ell GQ^*\right]^{1/2}.
\enge
\noi There are two horizons given by Eq.(\ref{109}) and the charge $Q^*$ is constrained by the real-valued square-root in the same
equation. In the next Section we will see that, in a braneworld scenario
Kerr-Newman BHs presents four horizons.

\section{Realistic Kerr-Newman mini-BHs at LHC}

Our assumption is now to consider a realistic mini-BH to be produced in LHC, which is  unlikely static, and the realistic conception
 is to proceed by considering a spinning,  charged
 Kerr-Newman mini-BH. It has been demonstrated \cite{jon} that half a mini-BH mass is emitted when it is
 highly rotating, confirming it is of primordial importance to take into account the angular momentum of BHs.
 The Kerr-Newman metric, in Boyer-Lindquist coordinates, describing the neighborhood of a spherical rotating BH with mass $M$, angular momentum $J$ and charge $Q$,
is given by 
\bege\label{eve} g_{\mu\nu}^{\rm K-N} = \begin{pmatrix}
\gamma/\rho^2&0&0&-\omega\be^2 + \alpha\\
0&\rho^2/\Delta&0&0\\
0&0&\rho^2&0\\
-\omega\be^2 + \alpha&0&0&\be^2
\end{pmatrix}\enge\noi where \beq\label{132}
\Delta &=& r^2 + \frac{a^2}{c^2} - 2\frac{GM}{c^2}r + Q^{2},\quad \rho^2 = r^2 + \frac{a^2}{c^2}\cos^2\theta,\nonumber\\
\Sigma^2 &=& \left(r^2 + \frac{a^2}{c^2}\right)^2 - \frac{a^2}{c^2}\Delta\sin^2\theta,\quad \be = \frac{\Sigma}{\rho}\sin\theta,\nonumber\\
\omega &=& \frac{2aGMr}{c^3\Sigma^2},\quad \alpha = \frac{a \sin^2 \theta Q^{2}}{c \rho^2}, \nonumber\\
\gamma &=& \frac{a^2\sin^2 \theta}{c^2} - \Delta.\eeq
The spinning parameter $a$ is defined as $a = \frac{J}{Mc}$.
In order to write the Kerr-Newman metric in a diagonal form, when we solve the characteristic eigenvalue equation associated with 
Eq.(\ref{eve}), the eigenvalues are given by 
\beq
\lambda_2 &=&\frac{\rho^2}{\Delta},\qquad\;
\lambda_3 = \rho^2,\nonumber\\
\lambda_{1,4} &=& \me\left[\right.\gamma/\rho^2 +  \be^2 \nonumber\\
&\pm& \left(\gamma^2/\rho^4 + 4(\alpha - \omega \be^2) - 2\beta^2\gamma/\rho^2 + \be^4\right)^{1/2}\left.\right]. \nonumber
\eeq\noi 
Now we must impose 
a condition --- arising when the eigenvalue characteristic equation 
is solved --- for
\emph{real} eigenvalues, i.e., 
\bege
\gamma^2/\rho^4 + 4(\alpha - \omega \be^2) - 2\beta^2\gamma/\rho^2 + \be^4\geq 0,
\enge\noi from which the Kerr-Newman metric is given in a diagonal form: 
\beq\label{kerr}
g^{\rm K-N} &=& g_{\mu\nu}^{\rm K-N}dx^\mu dx^\nu\nonumber\\ 
 &=& \lambda_1 {dt^{\prime}}^2 + \frac{dr^2}{\Delta/\rho^2} + \rho^2 d\theta^2 + \lambda^4d{\phi^\prime}^2.\label{metricap}
\eeq\noi Here $d\phi^\prime$ and $dt^\prime$ are 1-form fields on the 3-brane respectively related to $d\phi$ and $dt$ 
by the new eigenvectors in the associated directions defined by the eigenvalue equation associated with Eq.(\ref{eve}).

In order to obtain the correction of the Kerr-Newman radii, by braneworld effects, we follow the idea presented in Eq.(\ref{h}).
Besides, it is well known that a particular way to express the vacuum field equations on the brane is \cite{Maartens}
\begin{equation}\label{ansa1}
E_{\mu\nu} = - R_{\mu\nu},\quad R_\mu^{\;\;\mu} = 0 = E_\mu^{\;\;\mu},\quad  \nabla^\nu E_{\mu\nu} = 0.
\end{equation}
\noindent
Defining $\xi(r)$ as the deviation from a Kerr-Newman form  $\Delta/\rho^2$ --- 
the term in the denominator of $dr^2$ in Eq.(\ref{metricap}) --- we can now \emph{solve} Eqs. (\ref{ansa1}), via Eqs.(\ref{ewe}),
 for the particular case 
where $\mu=\nu=0$, obtaining 
\bege\label{133}
\xi(r) = \frac{2G\ell Q^*}{c^2r(r + a/c)}
\enge\noi as one solution, in such a way that in a static limit ($a\rightarrow 0$) the correction in the gravitational potential 
satisfies Eq.(\ref{potential2}). 
 Imposing that  the charge in these mini-BHs are induced \emph{a priori} by bulk effects, 
we can define $Q = GQ^*/c^2$. 
Now, the corrections ${R_{\rm K-Nbrane}}$ in the Kerr-Newman radii are obtained via the deviated Kerr-Newman form, as
\bege
 \frac{\Delta}{\rho^2} + \xi(R_{\rm K-Nbrane}) = 0.
\enge\noi By expanding the expression above and using Eqs.(\ref{132}) and (\ref{133}) 
we obtain 
\beq\label{cin}
R_{\rm K-Nbrane}^4 &&+ \Gamma_3R_{\rm K-Nbrane}^3 + \Gamma_2R_{\rm K-Nbrane}^2 \nonumber\\&& + \Gamma_1R_{\rm K-Nbrane} +\Gamma_0 = 0,
\eeq\noi 
where
\beq
&&\Gamma_3 = a/c - 2GM/c^2, \nonumber\\
&&\Gamma_2 = a^2/c^2 + 2\ell Q - 2aGM/c^3 + Q^2,\nonumber\\
&&\Gamma_1 = a^3/c^3 + aQ^2/c, \nonumber\\
&&\Gamma_0 = 2G\ell Q^* a^2\cos^2 \theta/c^2.
\eeq

\begin{figure}[H]
\centering
\includegraphics[width=7.0cm]{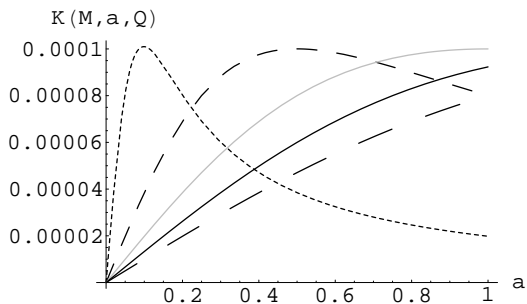}
  \caption{\small Graphic of the brane effect-corrected normalized inner horizon  $K(M,a,Q)$ $\times$ $ a $ for different values 
of $Q^*$. For the very short dashed line: $Q^* = -0.1$; for the short dashed line: $Q^* = -0.5$; for the full gray  $Q^* = -1$; for the full black line:  $Q^* = -1.5$; and for the long dashed line:  $Q^* = -2$.}
\label{n1}
\end{figure}

\begin{figure}[H]
\centering
\includegraphics[width=7.0cm]{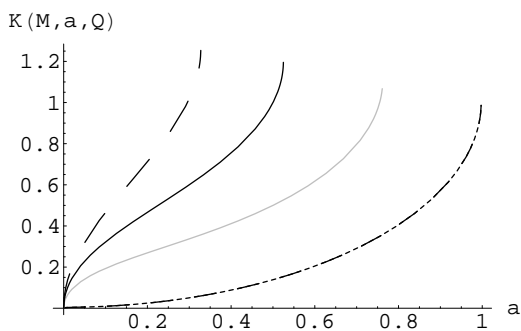}
  \caption{\small Graphic of the brane effect-corrected normalized horizon $K(M,a,Q)$ $\times$ $ a $ for different values 
of $Q^*$. These are solutions for a intermediate horizon. For the very short dashed line: $Q^* = -0.1$; for the short dashed
 line: $Q^* = -0.5$; for the full gray  $Q^* = -1$; for the full black line:  $Q^* = -1.5$; and for the long dashed line: 
 $Q^* = -2$. In this case, the spin of the mini-BH is constrained by the electric and tidal charge. The maximum $a$
 obtained for each graphic is: $a = 0.32806$ for $Q^*=-2$, $a = 0.52572$ for $Q^*=-1.5$,  $a = 0.76550$ for $Q^*=-1$,
 $a = 0.93842$ for $Q^*=-0.5$ and $a=0.99753$ for $Q^*=-0.1$ (which practically coincides with the $Q^*=-0.5$ case). 
In these respective values for $a$, the derivative of  $K(M,a,Q)$ related to $ a $ tends to infinity.}
\label{n2}
\end{figure}

\begin{figure}[H]
\centering
\includegraphics[width=7.0cm]{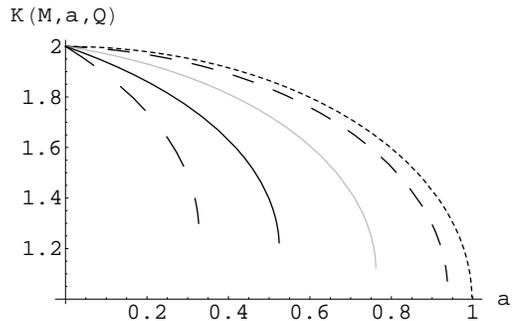}
  \caption{\small Graphic of the brane effect-corrected Kerr-Newman outer horizon in the form 
$c^2R_{K-N}/GM$ $\times$ $ a $ for different values 
of $Q^*$. For the very short dashed line: $Q^* = -0.1$; for the short dashed line: $Q^* = -0.5$; for the full gray  $Q^* = -1$; for the full black line:  $Q^* = -1.5$; and for the long dashed line:  $Q^* = -2$. In this case, the spin of the BH is constrained by the electric and tidal charge. The maximum $a$ obtained for each graphic is: $a = 0.32791$ for $Q^*=-2$, $a = 0.52766$ for $Q^*=-1.5$,  $a = 0.76585$ for $Q^*=-1$, $a = 0.95826$ for $Q^*=-0.5$ and $a=0.99866$ for $Q^*=-0.1$.}
\label{n3}
\end{figure}

\begin{figure}[H]
\centering
\includegraphics[width=8.4cm]{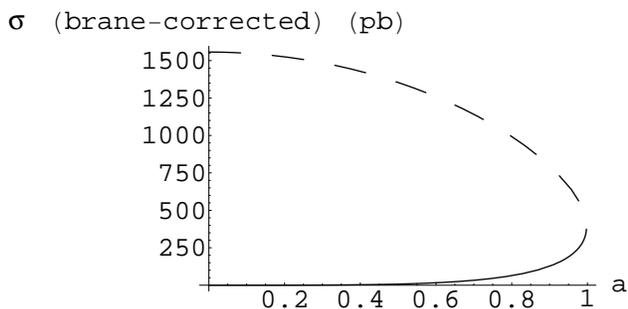}
  \caption{\small Brane effect-corrected Kerr-Newman cross section $\sigma$ for the inner horizon 
(full line) and for the outer horizon (dashed line) for various angular momenta $a$. Significant contributions by the inner
 horizon just happen when $a \rightarrow 1$. Thus, the effective contribution for the total cross section is only given by the outer
 horizon. This graphic presents an example for $n=3$ extra-dimensional ADD model, $M_P = 1$ TeV, $M_{BH}=5$ TeV, $Q^* = -0.1  M_{BH}$.}
\label{n4}
\end{figure}

A  mini-BH not corrected by extra-dimensional tidal charge effects can be simply  described by a Kerr mini-BH: the horizon
 radius comes from the expression obtained by making $Q = 0$ in Eq.(\ref{kerr}). Using natural 
units ($c=1$, $G=1$),
% and making the mass $M=1$, 
by solving Eq.(\ref{cin}) we find results for the normalized brane-corrected 
Kerr-Newman radius $K(M,a,Q)$. In what follows, this normalized horizon will be used to make corrections 
to a conventional Schwarzschild radius $R_S$, in order to find a Kerr-Newman BH by
$R_{\rm K-Nbrane} = \me K(M,a,Q)R_S$. 
Eq.(\ref{cin}) gives four solutions for $R_{\rm K-Nbrane}$ and we depict below three graphics 
respectively representing the inner $R_{\rm K-Nbrane}$ horizon corrections, 
  intermediary $R_{\rm K-Nbrane}$ radius corrections, and 
the outer $R_{\rm K-Nbrane}$ horizon corrections, 
 as functions of the spinning parameter $a$, for specific values 
of the charge $Q^*$.
Solutions for the normalized horizon $K(M,a,Q)$ with respect to the spinning parameter factor $a$ 
is presented in Figures (\ref{n1}), (\ref{n2}),
 and (\ref{n3}), respectively for inner, intermediate, and outer Kerr-Newman horizons. 
These corrections can be applied to the Schwarzschild radius $R_S$ of a Myers-Perry $(4+n)$-dimensional BH \cite{myers,argyres,dim1}, 
resulting in
\beq\label{myers}
R_{\rm K-Nbrane} &=& \me K(M,a,Q)R_S \nonumber\\
&=& \frac{K(M,a,Q)}{2\sqrt{\pi}M_{P(4+n)}}\left[\frac{\tt M_{BH}}{M_{P(4+n)}}\left(\frac{8 \Gamma \left(\frac{n+3}{2}\right)}{n+2}\right)\right]^{\frac{1}{n+1}}
\eeq
\noi where $\Gamma(m)$ denotes the usual Euler Gamma function, $M_{P(4+n)}$ is the $(4+n)$-dimensional Planck mass and $M_{BH}$ 
is the mini-BH mass. Both $M_{P(4+n)}$ and $M_{BH}$ are given in TeV. Expression (\ref{myers}) is normalized ($c = 1$, $\hbar = 1$) and to obtain the value of $R_{\rm K-Nbrane}$ in meters it is necessary to multiply the expression by $\hbar c = 1.973 \times 10^{-13}$ MeV m. The three solutions for $K(M,a,Q)$, represented by the graphics of Figs. (\ref{n1}), (\ref{n2}), and (\ref{n3})  generate three horizons, which could be related to quantum BH theories \cite{bekenstein}. Only the outer horizon expressed by Fig. (\ref{n3}) contributes to the total cross section (see Fig. (\ref{n4})). 

Denoting by $M_*$ the strength of higher-dimensional gravity, $M_P\approx$ $ 10^{16}$ TeV 
sets the strength of 4-dimensional gravity, and $V^n$ the volume of the higher-dimensional space, 
the 3-space-dimensional Newton's force law is recovered by the identification \cite{dim1,Maartens,land,land1}
\bege
M_P^2 = M_*^{n+2} V^n.\nonumber
\enge
\noindent Important differences concerning the number of extra dimensions 
in ADD and Randall-Sundrum formalism resides in the last expression. 
In the one hand, if there were only one large extra dimension in ADD model,
gravity measured at, e.g., the solar system, would have a 5-dimensional non-observed behavior. With $n=2$ extra dimensions the size of extra 
dimensions would be a tenth of a millimeter. On the other hand, the Randall-Sundrum model is effectively 5-dimensional, 
with one extra warped dimension. In the graphics below we present our results for certain number of extra dimensions, 
and it is implicit that, when we consider $n=1$ extra dimension it is associated with Randall-Sundrum models, while
for $n=2,\ldots,6$ extra dimensions our results represent ADD models. 

Considering two partons with the center of mass energy $\sqrt{\tilde{s}}= M$ moving in opposite directions, 
semi-classical reasoning suggests that the impact parameter is less than the Schwarzschild radius --- in the case of
an extremal Kerr-Newman mini-BH --- or less than the outer Kerr-Newman horizon --- when the Kerr-Newman mini-BH \emph{is not} extremal ---
 a mini-BH with the mass $M$ arises. For the Schwarzschild BH particular case, see \cite{dim1,land,land1}. 
Parton distribution functions at LHC gives the total cross section $\sigma$ for production of BHs with $M_{BH} > M_{P}$
in the range  1 pb  $< \sigma <$  15 pb,   for 1 TeV  $ < M_P <$  5 TeV and it varies around 10\% for 
$n$ --- the number of extra dimensions in ADD models ---
between 2 and 7 \cite{land,land1}. 

\begin{figure}[H]
\centering
\includegraphics[width=8.4cm]{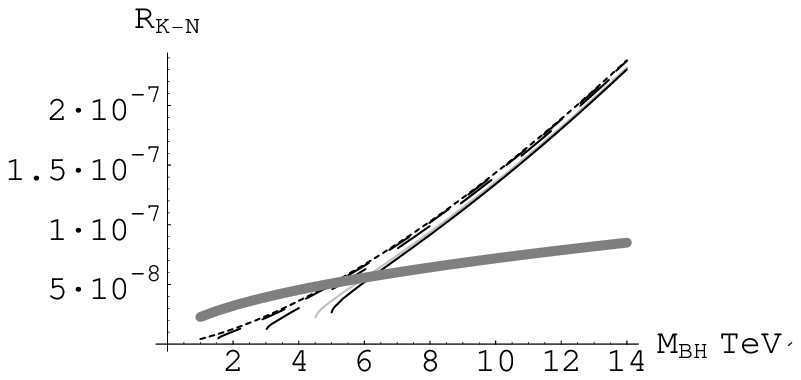}
\includegraphics[width=8.4cm]{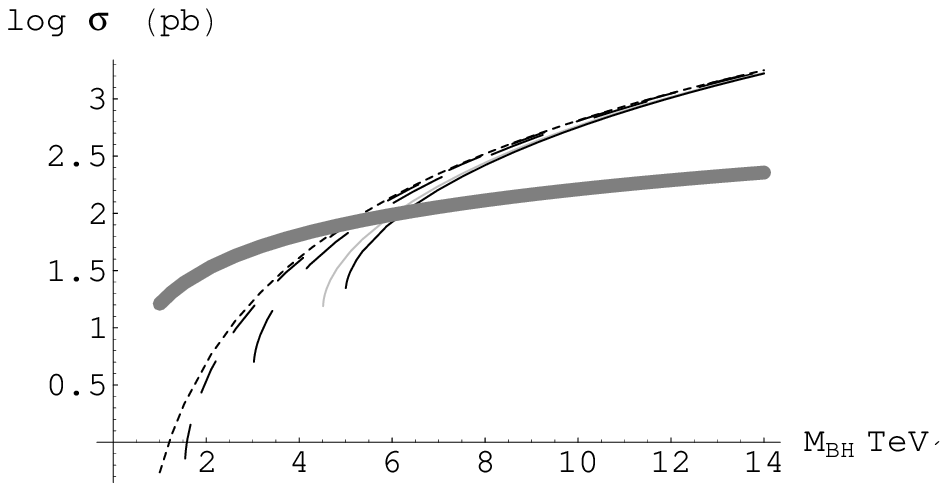}  \caption{\small Graphics of the brane effect-corrected Kerr-Newman horizon R$_{{\rm K-N}}$ 
of the produced mini-BH and its associated cross section, 
for various momenta $a$ versus BH masses, and its associated  parton cross section 
for Randall-Sundrum braneworld model, 
where there is one warped extra  dimension. 
Here we present an example for $M_P = 4$ TeV, and $Q^* = -0.1 M_{BH}$, with short dashed line for $a=0.1$, 
dashed line for $a=0.3$, long dashed line for $a=0.6$, gray full line for $a=0.9$,
full black line for $a=0.997$.}
\label{n5}
\end{figure}

\begin{figure}[H]
\begin{center}
\includegraphics[width=8.4cm]{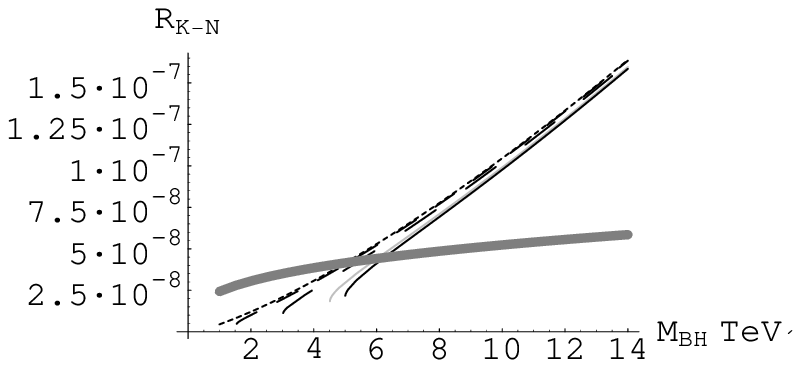}
\includegraphics[width=8.4cm]{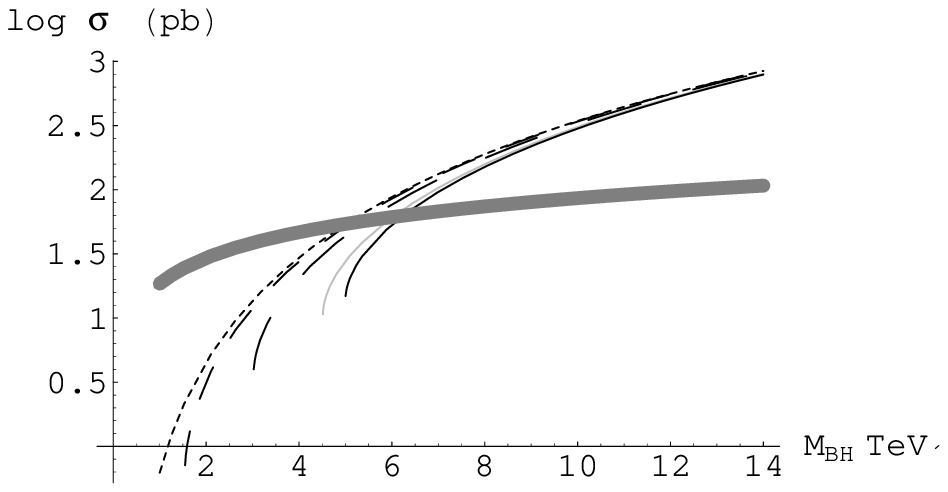}
  \caption{\small Graphics of the brane effect-corrected Kerr-Newman radius and its associated  parton cross section for $n = 2$ ADD model 
for some BH masses, with $M_P = 1$ TeV, and for various angular momenta: short dashed line for $a=0.1$, dashed line for $a=0.3$, long dashed line for $a=0.6$, gray full line for $a=0.9$,
full black line for $a=0.997$.}
\label{rkncsd2a}
\end{center}
\end{figure}

\begin{figure}[H]
\centering
\includegraphics[width=8.4cm]{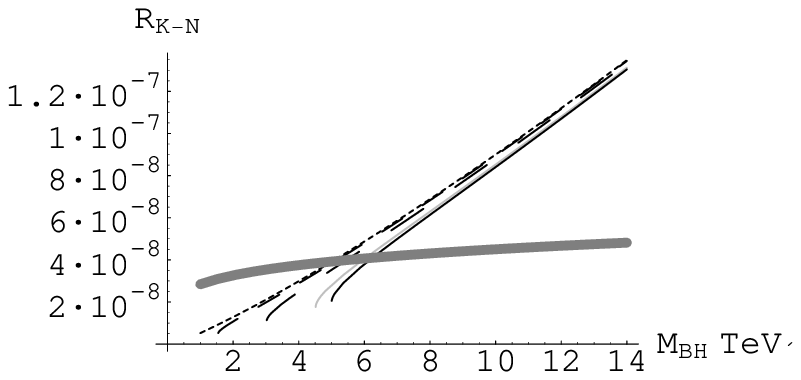}
\includegraphics[width=8.4cm]{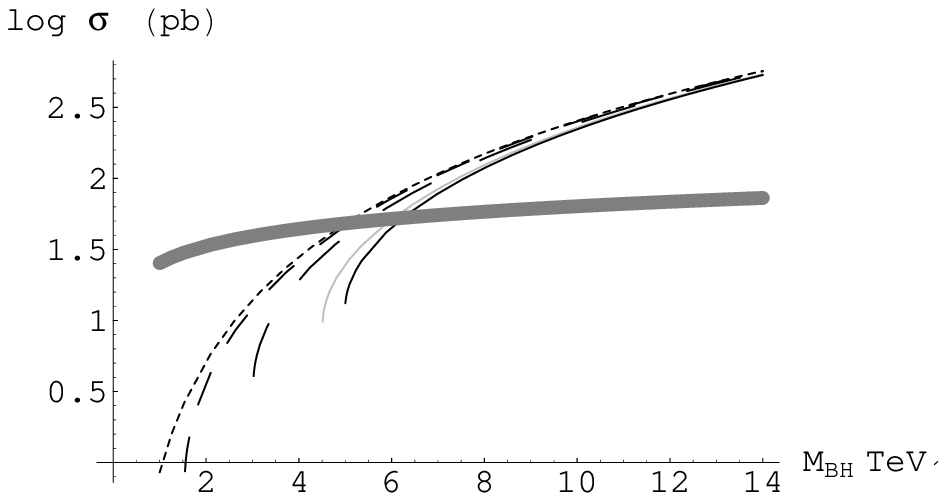}
\caption{\small Graphics of the brane effect-corrected Kerr-Newman outer horizon and 
its associated  parton cross section in $n=4$ ADD model for some mini-BH masses, 
with $M_P = 1$ TeV, and for various angular momenta: short dashed line for $a=0.1$, dashed line for $a=0.3$, 
long dashed line for $a=0.6$, gray full line for $a=0.9$,
full black line for $a=0.997$. }
\label{rkncsd4a}
\end{figure}

\begin{figure}[h]
\centering
\includegraphics[width=8.4cm]{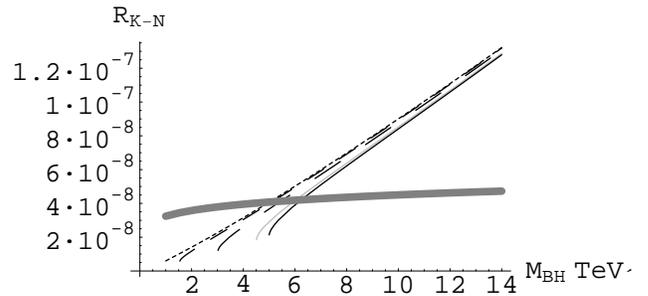}
\caption{\small Graphic of the brane effect-corrected Kerr-Newman outer horizon for extra-dimensional  
$n = 6$ ADD model for some BH masses, with $M_P = 1$ TeV, and for various angular momenta: short 
dashed line for $a=0.1$, dashed line for $a=0.3$, long dashed line for $a=0.6$, gray full line for $a=0.9$,
full black line for $a=0.997$.}
\label{rkncsd6a}
\end{figure}

\begin{figure}[H]
\centering
\includegraphics[width=8.4cm]{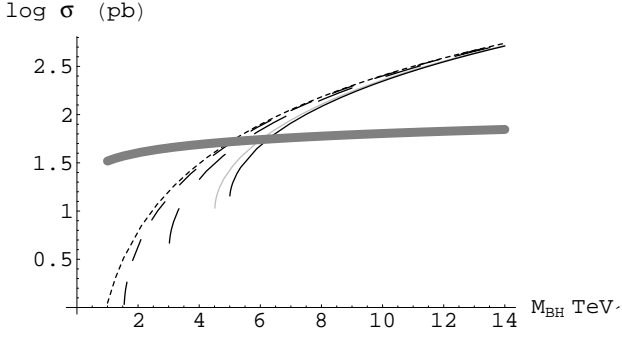}
\caption{\small Cross sections  for $n=6$ extra dimensions in ADD model, 
 short dashed line for $a=0.1$, dashed line for $a=0.3$, 
long dashed line for $a=0.6$, gray full line for $a=0.9$,
full black line for $a=0.997$.}
\label{rkncsd6b}
\end{figure}

\begin{figure}[H]
\centering
\includegraphics[width=8.4cm]{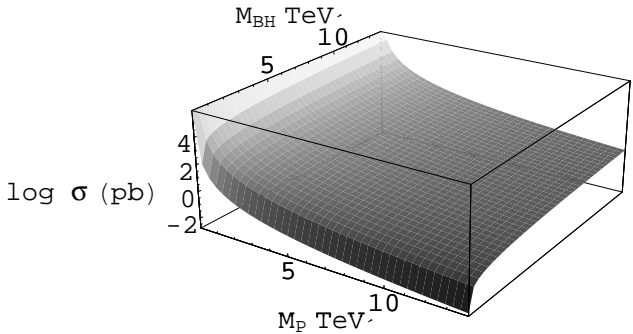}
  \caption{\small Graphic of parton cross section in picobarn for various values of Planck mass $M_{P}$ and BH mass $M_{BH}$ \emph{for 
the Schwarzschild Myers-Perry model}. The mini-BH formed by the scattering has $a = 0.5$ and is  charged ($Q^* = -1M_{BH}$), 
for the Randall-Sundrum braneworld model.}
\label{RS22}
\end{figure}

\begin{figure}[H]
\centering
\includegraphics[width=8.4cm]{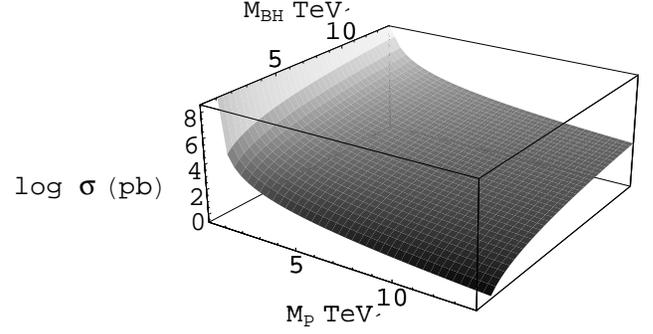}
  \caption{\small Graphic of parton cross section in picobarn for various values of Planck mass $M_{P}$ and BH mass $M_{BH}$. The mini-BH formed by the scattering has $a = 0.5$ and is  charged with $Q^* = -1M_{BH}$, for the Randall-Sundrum braneworld model.}
\label{RS2}
\end{figure}

\begin{figure}[H]
\centering
\includegraphics[width=8.4cm]{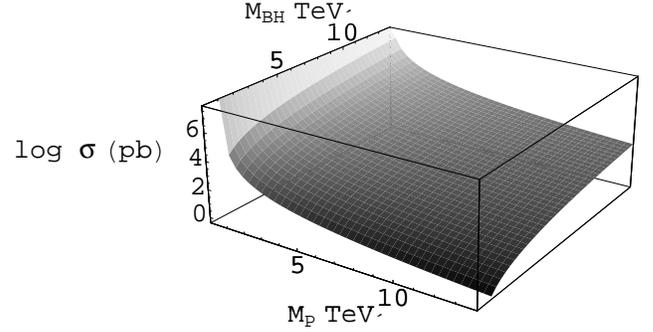}
\includegraphics[width=8.4cm]{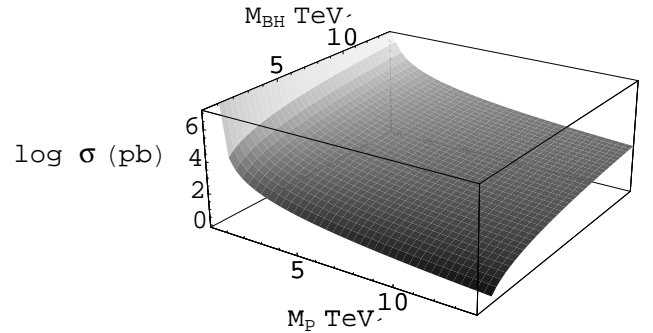}
\caption{\small Graphics of parton cross section in picobarn for various values of Planck mass $M_{P}$ and BH mass $M_{BH}$. The mini-BH formed by the scattering has $a = 0.5$ and is  charged with $Q^* = -1M_{BH}$, in respectively $n=2$ and $n=6$ 
extra dimensions in ADD model. }
\label{csmbmpA}
\end{figure}

\begin{figure}[H]
\centering
\includegraphics[width=8.4cm]{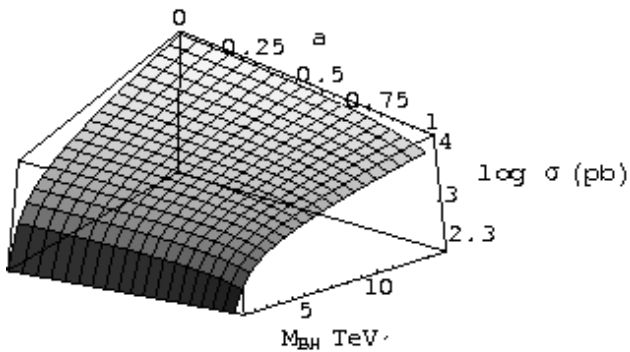}
  \caption{\small Graphic of parton cross section in picobarn for various values of spinning parameter $a$ 
and BH mass $M_{BH}$. The mini-BH formed by the scattering  is  charged with $Q^* = -1M_{BH}$. In this case we consider the Randall-Sundrum braneworld model.}
\label{RS3}
\end{figure}

\begin{figure}[H]
\centering
\includegraphics[width=8.4cm]{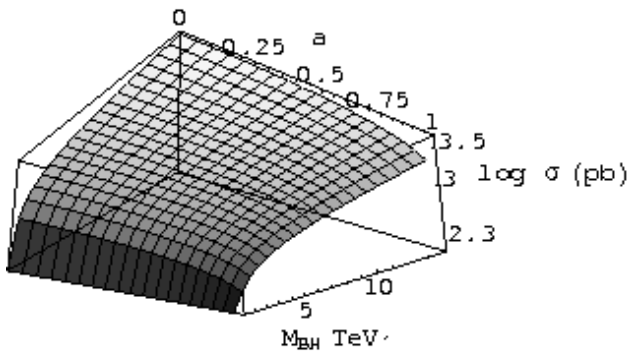}
\caption{\small Graphic of parton cross section in picobarn for various values of $a$ and the BH mass $M_{BH}$. 
Note the interesting effect caused by the spinning 
parameter $a$ in the cross section. Here we consider $n=2$ extra-dimensional ADD model, 
the mini-BH is electrically and tidally charged by the bulk with $Q^* = -0.1M_{BH}$.}
\label{csmbaA}
\end{figure}

\begin{figure}[H]
\centering
\includegraphics[width=8.4cm]{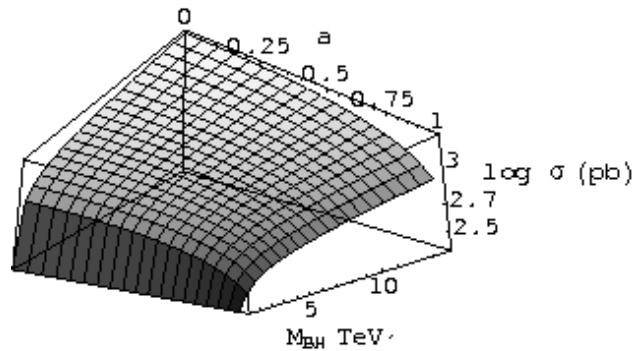}
\caption{\small Graphic of parton cross section in picobarn for various values of $a$ and the BH mass $M_{BH}$. 
Note the interesting effect caused by the spinning 
parameter $a$ in the cross section. In this case $n=6$ extra-dimensional ADD model the bending is more intense. 
The mini-BH has $Q^* = -0.1M_{BH}$.}
\label{csmbaB}
\end{figure}

By elementary geometrical arguments,  the total cross section is given by \cite{empa,myers} 
\beq
\sigma(M_{BH}) && \approx \pi R^2_{\rm K-Nbrane}\nonumber \\
&&= \frac{K^2(M,a,Q)}{4M^2_{P(4+n)}}\left[\frac{M_{BH}}{M_{P(4+n)}}\left(\frac{8 \Gamma \left(\frac{n+3}{2}\right)}{n+2}\right)\right]^{\frac{2}{n+1}}.\nonumber
\eeq
Now, in Figs. (\ref{n5}, \ref{rkncsd2a}, \ref{rkncsd4a}, \ref{rkncsd6a}, \ref{rkncsd6b}) 
we illustrate the brane effects 
in Kerr-Newmman mini-BHs --- for different values of charge --- as functions of mini-BH mass, Planck mass and spinning parameter
(respectively in Randall-Sundrum and $n=2,4$ and 6 extra-dimensional ADD models). 
\emph{In these graphics the wide gray line indicates the standard Myers-Perry approach for Schwarzschild mini-BHs}. We show in this way 
a more realistic and precise prediction concerning mini-BH production in LHC, since our approach does not neglect spin and charge effects.
In Fig. (\ref{RS22}) we show the parton cross section for the case of Schwarzschild  mini-BH Myers-Perry 
production within LHC range of masses, and in 
Fig. (\ref{RS2}) we show the parton cross section for the case of production of a spinning charged mini-BH within LHC range of
 masses in braneworld Randall-Sundrum 
scenario. We present the same graphics for $n=2$ and $n=6$ extra dimensions, respectively in Fig. (\ref{csmbmpA}), 
and 
in these cases, we obtain an attenuation in the cross section values when the BH mass increases.
 In Fig. (\ref{RS3}) we show the parton cross section for the case of production of a spinning mini-BH within the
 LHC range of masses, and for the case where the scenario is a Randall-Sundrum braneworld model; here  we show 
explicitly the effect of the spinning parameter $a$. We present the same graphic for higher dimensions in Figs.
 (\ref{csmbaA}) and (\ref{csmbaB}). It is notorious the interesting effect caused by the spinning parameter $a$ in the cross section. The 
higher the number of extra dimensions, for instance $n=6$ as illustrated in Fig.(\ref{csmbaB}), 
the faster the bending of the graphic is, as the spinning parameter $a\rightarrow$ 1.
\begin{figure}[H]
\centering
\includegraphics[width=8.4cm]{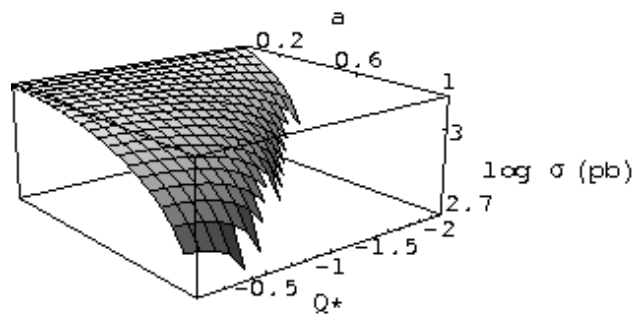}
\caption{\small Parton cross sections in picobarn for various values of $Q^*$ and $a$, 
explicitly showing  the influence in the scattering if one considers a realistic Kerr-Newman mini-BH. The charge constrains the spin.}
\label{csqa}
\end{figure}
\section{BH evaporation and Hawking radiation in Kerr-Newman mini-BHs}

 BHs 
can evaporate by emitting pairs of virtual photons at the event horizon, with one of the photon escaping the BH gravity \cite{haw}, 
and have black-body spectrum with the Hawking temperature $\frac{\hbar c}{4\pi k R_S}$. 
Hawking evaporation is significant only in the case of small BHs, where 
the tidal effect becomes so intense near the surface that the particle pairs produced by quantum 
vacuum fluctuations may be broken, one particle falling into the BH and the other being projected outwards.

BH production in LHC is expected immediately above the $M_P$ threshold, since at energies below 
$M_P$ other quantum effects overcome BH production. In \cite{land,land1} it was shown how the relationship between
$M_{BH}$ and Hawking temperature can unravel Planck mass and the dimensionality $n$ of extra dimensions, independently of the 
geometry of extra dimensions.
Although when a BH event horizon is observed by its Hawking radiation it looks like a fuzzy sphere (see \cite{fuzzy})
--- and in the classical limit the event horizon looks locally like a non-commutative plane with
non-commutative parameter dictated by the Planck length --- in particular at 
$M_P$ energies BHs are quantum objects and here we use as in \cite{land1} semi-classical 
arguments. Besides, mini-BHs produced at LHC would be light and 
 extremely hot, with a Hawking temperature $T_H \approx 100$ GeV, 
and evaporate almost instantaneously, mainly via Hawking radiation.

Evaporation of a BH in $n$ large extra dimensions occurs at Hawking temperature $T_H$, given in our model by
\begin{equation}
T_H  = \frac{2M_P}{K(M,a,Q)} \left(\frac{M_{BH}}{M_P}\frac{n+2}{8\Gamma(\frac{n+3}{2})}\right)^\frac{1}{n+1}\frac{n+1}{4\sqrt{\pi}}.
\end{equation}
\noi We depict below $T_H$ for different values of Kerr-Newman mini-BH charges: 
\begin{figure}[H]
\centering
\includegraphics[width=8.4cm]{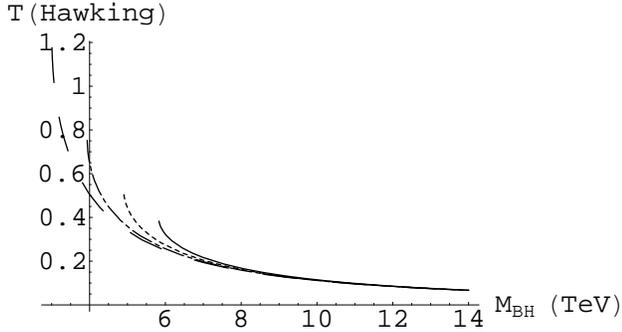}
\caption{\small Range of Hawking temperature, in TeV, produced in LHC by a charged spinning BH evaporation versus BH mass interval. The range covers some values of charge for a fixed angular momentum: full line: $Q^* = -2M$, dotted line: $Q^* = -1.5M$, dashed-dotted: $Q^* = -M$, dashed: $Q^* = -0.5M$, all for spinning parameter $a = 0.5$. This means the BH charge is a important parameter in the determination of exact BH decay. Here, the Planck mass is $4$ TeV and $n=6$ for Randall-Sundrum braneworld model.}
\label{hawking}
\end{figure}

In this case, as the charge $Q^*$ increases, the Hawking temperature
increases as well. 
On the other hand, as the mini-BH mass increases,
$T_H$ decreases, and for all values of $Q^*$, Hawking temperature $T_H$ tends to the same value as $M_{BH}$ is $\sim 10$ TeV.

\begin{figure}[H]
\centering
\includegraphics[width=8.4cm]{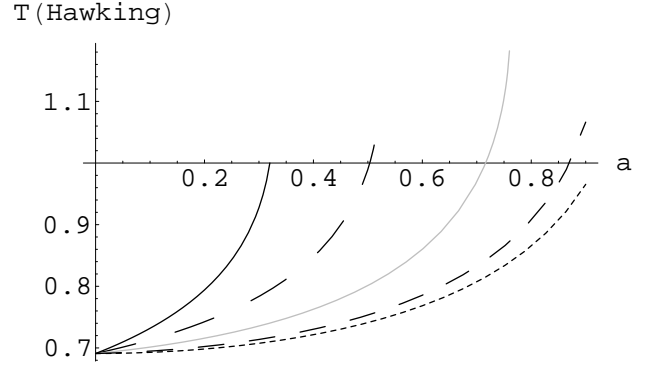}
\caption{\small Range of Hawking radiation, in TeV, produced in LHC by a charged spinning BH evaporation versus BH spinning parameter
 $a$, for a fixed BH mass (1 TeV) for Randall-Sundrum braneworld model. Here we present an example for $M_P = 4$ TeV. For the very short dashed line: $Q^* = -0.1$; for the short dashed
 line: $Q^* = -0.5$; for the full gray  $Q^* = -1$; for the full black line:  $Q^* = -1.5$; and for the long dashed line:  
$Q^* = -2$. In this case, the spinning of the BH is constrained by the electric and tidal charge.}
\label{hawking1}
\end{figure}

It is notorious by the graphic above that as the spin parameter $a$ increases
 $T_H$ increases as well.

\begin{figure}[H]
\centering
\includegraphics[width=8.4cm]{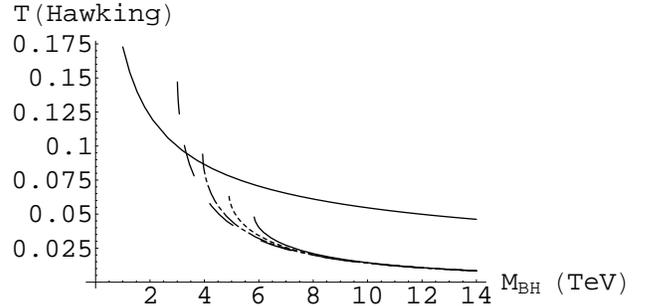}
\caption{\small Range of Hawking radiation, in TeV, produced in LHC by a charged spinning BH evaporation versus BH spinning parameter
 $a$, for a fixed BH mass (1 TeV) for Randall-Sundrum braneworld model, where there is one extra 
warped dimension. Here we present an example for Planck mass $M_P = 1$ TeV. For the very short dashed line: $Q^* = -0.1$; for the short dashed
 line: $Q^* = -0.5$; for the full gray  $Q^* = -1$; for the full black line:  $Q^* = -1.5$; and for the long dashed line:  
$Q^* = -2$. In this case, the spinning of the BH is constrained by the electric and tidal charge.}
\label{hawking2}
\end{figure}

In this case Hawking temperature is about $\sim 10^1-10^2$ GeV.
The superior line indicates the graphic for a 
Schwarzschild BH Myers-Perry model \cite{dim1,land}. 

\section{Concluding Remarks and Outlooks}

By solving Einstein equations on the brane we have found an explicit form for the perturbation of a 
Kerr-Newman metric singularity, corresponding to a Kerr-Newman BH. Such perturbation gives rise to 
more realistic calculations and predictions involving mini-BH production at LHC, since we concern 
 braneworld corrections as they would appear in the measurement of cross sections, Hawking temperature and some
other physical observations related to the mini-BH production at LHC.
Mini-BHs are a natural choice for testing braneworld formalisms: 
the lighter the BH the greater its surface gravity is,  and consequently  mini-BHs
are more sensitive on braneworld physical effects, like, e.g, Hawking evaporation in the context of Randall-Sundrum 
and ADD models. Also, 
%As the real $D$-dimensional fundamental Planck scale is given by a TeV order of magnitude, a braneworld scenario is effectively a natural choice
%to the hierarchy problem solution. In fact, large extra dimensions  provide an alternative to SUSY 
%in addressing the hierarchy problem and in this case gravity could be stronger than we think,  and the Planck scale 
%may be within reach of the LHC. 
%A spectacular consequence of such a model is the possibility of being able to produce  mini-BHs 
%in the next generation of particle colliders. 
the 14 TeV maximum energy of LHC could allow it to become a mini-BH factory with a production rate as high as about one per second, and  
%The hadron to lepton ratio of the decay products is roughly expected to be 5:1. 
%It is usual to define a parameter $\alpha$ in order to exploit the deviation of the uncertainty principle, as \cite{lhc1}
%\begin{equation}
%\Delta x = \frac{\hbar}{\Delta p}\left[1 + \alpha\ell_{\rm Planck}\left(\frac{\Delta p}{\hbar}\right)^2\right],
%\end{equation}\noindent where $\ell_{\rm Planck}$ denotes Planck length.
%and it is well known in the case where $M_{BH}$ = 12 TeV that $\alpha = 0$ --- for the General Uncertainty Principle --- 
%and the BH evaporates into five quarks, one charged lepton, and one gluon, with the charged leptons accounting for about 15 \% of the emission. As the parameter $\alpha$ decreases, the leptonic component becomes negligible \cite{dim1}. 
after formation,  mini-BHs are expected to lose the information associated with multipole and angular momenta,
 to decay via Hawking radiation, and 
eventually either disappear completely or leave Planck-sized remnant \cite{dim1,duffy,rem1,ashu} around 
Planck masses   $M_{P}=1.6$ TeV for $D=6$ and $M_{P}= 0.25$  TeV for $D=10$ \cite{lhc1}.

%The possible presence of extra dimensions would be doubly beneficial for the production of mini-BHs. 
 %Schwarzschild radius is significantly increased in a non-extremal RN BH by Eq.(\ref{109}) \cite{Maartens}. 
%The resulting mini-BHs have radii
% that are much smaller --- $\sim$ $10^{-4} fm$  in the case of those expected from the LHC --- than the size of 
%extra dimensions, and that they can therefore be considered as 
%totally embedded in a $D$-dimensional space. 

Figures (\ref{n1}), (\ref{n2}), and (\ref{n3}) in Section II show the corrections of Schwarzschild radius as functions 
of the spin parameter $a$, for different values of the charge $Q^*$. Such brane effect-corrections arise 
immediately from the solutions of Einstein equations on the brane and they can describe more realistic 
predictions concerning mini-BHs features at LHC. 

%We have explored the braneworld Randall-Sundrum model, but on a 10-dimensional fiber bundle Horava-Witten braneworld 
%scenario could bring a more promising theoretical paradigm.
%It is well known that, as the particles residing in the brane greatly outnumber those living in the bulk 
%--- which are essentially gravitons ---
% the mini-BH evaporates into particles of the Standard Model, such as photons,
% electrons, and quarks, with energies ranging from 80 GeV down.
Brane effect-corrections for mini-BH cross sections at LHC are shown in Figure (\ref{n4}) for $n=3$ extra dimensions in ADD model.
Figures (\ref{n5}),  (\ref{rkncsd2a}), (\ref{rkncsd4a}), (\ref{rkncsd6a}), and 
(\ref{rkncsd6b})
 also show the increment of the brane effect-corrected Kerr-Newman BH horizons and the respective cross section
as functions of the Kerr-Newman BHs, for different values of charge, as functions of BH and Planck masses, 
and spinning parameter, respectively for the Randall-Sundrum model and the $n=2$, $n=4$, and $n=6$ extra-dimensional ADD models.
In all these graphics the wide gray line indicates the standard Myers-Perry approach for Schwarzschild BHs,
and it is clear that our results coming from Kerr-Newman analysis predict more accurately
the results obtained from the static Schwarzschild Myers-Perry approach.

Figures (\ref{RS22}), (\ref{RS2}), (\ref{csmbmpA}) definitively express the brane corrections in mini-BH cross sections 
as functions of BH mass  and the effective Planck mass in the range 1-10 TeV of LHC, respectively for Schwarzschild --- in Myers-Perry
model ---
and Kerr-Newman mini-BHs for various
values of extra dimensions both in Randall-Sundrum and ADD models. Figures 
(\ref{RS3}), (\ref{csmbaA}), and (\ref{csmbaB}) show the brane corrections in mini-BH cross sections 
as functions of the spinning parameter $a$ and the mini-BH mass, in the range 1-10 TeV of LHC, respectively
for various values of $n$ extra dimensions in Randall-Sundrum ($n=1$), and ADD models (the particular cases where $n=2$ and $n=6$). 
Our results show prominent effects of higher number of extra dimensions in cross sections and Hawking temperature associated 
with Kerr-Newman mini-BHs.
Fig.(\ref{csqa}) illustrate the brane-effect corrected cross sections as functions of $Q^*$ and the spinning parameter $a$, showing how 
the charge constrains the mini-BH spin.

Figures (\ref{hawking}), (\ref{hawking1}), and (\ref{hawking2}) illustrate Hawking temperature 
of a Kerr-Newman mini-BH respectively as function of BH mass and spinning parameter $a$, 
both in an extra-dimensional scenario.
Figure (\ref{hawking1}) show the braneworld Kerr-Newman mini-BH corrections
of Schwarzschild Myers-Perry model. Again, for different values of mini-BH charge, 
$T_H$ seems not to be sensitive to these values for higher values of mini-BH mass, above 10 TeV.
This shows that in the range 1-10 TeV of LHC resides our brane effect-corrections. 
Also, from Figs. (\ref{hawking}), (\ref{hawking1}), and (\ref{hawking2}), as the charge $Q^*$ increases, the Hawking temperature
increases as well. On the other hand, the greater the mini-BH mass,
the lower the $T_H$, and for all values of $Q^*$, Hawking temperature $T_H$ tends to the same value as $M_{BH}$ is $\sim 10$ TeV. 
This allows us
to follow \cite{land,land1} in a braneworld viewpoint, and to predict ---
from the graphic of Hawking temperature  $\times$ $M_{BH}$ --- the
 Planck mass and the number $n$ of extra dimensions in a more precise and realistic approach.
The fact that $T_H$ increases as the spinning parameter $a$ also increases shows that 
all more realistic brane effect-corrections obtained in this paper for cross sections, Hawking temperature, and Kerr-Newman horizons, 
can be extremely useful in an optimal use of the range 1-10 TeV of LHC. 

The question concerning where mini-BHs mostly radiate --- on the brane or on the bulk --- 
remains open, and its answer will allow, at least in principle, to detect most of the decay products in the case of radiation on the brane.
If BHs radiate mostly in the bulk, one has to take any
constraints coming from non-observation of such events with caution.
Due to non-zero impact parameter of initial particles, most of the
produced BHs will be highly rotating, and our model described in this paper 
describe such processes with more accuracy. Such BHs radiate
mostly in the bulk --- at least as long as they rotate fast --- as
first pointed out  in \cite{sto1,sto2,sto3}. 
The reason is that rotating BHs exhibit the
effect of superradiance, i.e., certain radiation modes
get significantly amplified taking away rotational energy of the BH \cite{s1,s2,sto4,sto5}. 
The model described here using Kerr-Newman electrically charged and rotating BHs 
gives rise to a more profound analysis of superradiance effect, which is highly spin-dependent and
radiation of higher spin particles is highly preferred --- for instance gravitons are
preferred to photons and neutrinos. While on a brane with (1+3)-dimensions there are two
degrees of freedom of graviton, the probability of a graviton emission is
a hundred times higher than the probability of a photon and neutrino emission
\cite{don1,don2}. This effect must be even stronger in the higher-dimensional case where you have more gravitational degrees of freedom
available. Also, superradiance exists in higher dimensions \cite{sto4,sto5}
and interactions between BHs and the brane \cite{sto6,sto7,sto8,sto9}
exist, in which BH lose some of its components
of angular momentum. Thus, after initial bulk domination, in final
stages, it is expected again that brane radiation dominates.
Also, the extra-dimensional geometrical scenario 
is fundamental for such analysis.
New analysis procedures involving superradiance and charged Kerr-Newman
mini-BHs are to be discussed in a forthcoming paper.

Besides, gravitational field equations on the brane can also be solved in the framework of 
ADM \cite{adm} formalism.  For instance, in \cite{goll} SMS effective gravitational field equation on the brane was recovered
 and it generalized  the off-brane equations due to both the acceleration of the  timelike 1-form fields, in the presence of
general bulk energy-momentum tensor. 

Finally, a mini-BH demands  a sufficiently
large entropy $S_{BH}$ \cite{king}, where 
the threshold mini-BH mass is much larger, and therefore the production
becomes much smaller at the LHC.  For example, to require
$S_{BH} > 25$  implies $M_{BH} > 5 M_P$
and if one assumes $M_P = 2$ TeV, then the threshold mini-BH mass is about 10 TeV, again
around the limit energy in LHC.

\section{Acknowledgment}
The authors are very grateful to Profs. Patricio A. Letelier, Dejan Stojkovic, Roberto Casadio, Ricardo Antonio Mosna
 and  Vitor Cardoso 
for important comments about this paper, 
and to Profs. Alikram Aliev and  Ignatios Antoniadis for pointing out some missing points 
in a previous version of this paper. The authors are also grateful to PRD
 referee for important  suggestions, in particular those in order to improve 
the  readability of the text.
Rold\~ao da Rocha thanks to Funda\c c\~ao de Amparo \`a Pesquisa
do Estado de S\~ao Paulo (FAPESP) for financial support and Carlos H. Coimbra-Ara\'ujo thanks to CAPES/Brasil for the financial support.

\end{document}